\documentclass[oribibl]{llncs}
\usepackage{version}
\excludeversion{ConfPaper}
\includeversion{TechnicalReport}
\usepackage{graphicx}
\usepackage{latexsym}
\usepackage{xspace}
\usepackage{url}

\newcommand{\ourModel}{NPC\xspace}

\newcommand{\mysubsubsection}[1]{\textbf{#1}}

\begin{document}

\title{The \ourModel Framework for Building Information Dissemination Networks}
% \title{An Unifying Framework for Information Dissemination, Retrieval, and Archiving}

\author{Lukas C.\ Faulstich}

\institute{Institut f\"ur Informatik,
  Universit\"at Leipzig\\ \email{faulstic@inf.fu-berlin.de} }
% \date{\today}

\maketitle

\begin{abstract}
  Numerous systems for dissemination, retrieval, and archiving of
  documents have been developed in the past. Those systems often focus
  on one of these aspects and are hard to extend and combine.
  Typically, the transmission protocols, query and filtering languages
  are fixed as well as the interfaces to other systems. We rather
  envisage the seamless establishment of networks among the providers,
  repositories and consumers of information, supporting information
  retrieval and dissemination while being highly interoperable and
  extensible.

  We propose a framework with a single event-based mechanism that
  unifies document storage, retrieval, and dissemination. This
  framework offers complete openness with respect to document and
  metadata formats, transmission protocols, and filtering
  mechanisms. It specifies a high-level building kit, by which
  arbitrary processors for document streams can be incorporated to
  support the retrieval, transformation, aggregation and
  disaggregation of documents. Using the same kit, interfaces for
  different transmission protocols can be added easily to enable the
  communication with various information sources and information
  consumers.
% 
%   Information dissemination networks within the Internet can be
%   established in various ways, using different approaches, protocols,
%   and information systems. Often such solutions have been developed
%   with a certain narrow application in mind and cannot be adapted
%   easily to support other more general information dissemination
%   tasks. Our vision is to enable common users to create networks for
%   disseminating, filtering, and transforming streams of documents just
%   like the Web enabled users to create static hypertext.

%   Existing information dissemination approaches prevent the
%   implementation of this vision by their lack of modularity,
%   interoperability, dynamic configurability, and by their commitment to
%   fixed filter languages, transport paradigms, document
%   manipulation facilities, and security policies. This makes it
%   difficult to combine technically heterogeneous information
%   dissemination systems and to establish and manage value-adding
%   information services on top of existing information sources or
%   services.

%   In this work we present the unified \ourModel framework that
%   provides a high-level building kit for a wide range of modular
%   information dissemination networks. Our framework interfaces well
%   with existing protocols and systems and can be implemented easily on
%   top of basic technology such as HTTP, file systems or RDBMS.
\end{abstract}

\section{Introduction}\label{sec:introduction}

The purpose of digital libraries is the archiving of electronic
documents and the support of their retrieval and dissemination. We
observe digital libraries as ``information dissemination networks''
(IDNs), the nodes of which are information providers, such as authors
and publishers, repositories (e.g.\ libraries, technical report
servers), and information consumers. The links among them may be
short-term, as in the case of a literature recherche, or long-term, as
for the subscriptions to alerting services and the mirroring of
repositories. Such networks must support various document and metadata
formats, transmission protocols, and heterogeneous query languages and
filters.

In this work, we present a unifying framework for the construction of
IDNs. It consists of a simple conceptual model and an event-based
mechanism for dissemination and retrieval of information. An
implementation of our ``\ourModel'' framework provides a building kit
consisting of \textbf{N}odes, \textbf{P}orts and \textbf{C}hannels.
%% </MYRA>

\begin{figure}[htb]
  \begin{center}
    \includegraphics*[width=0.66\textwidth]{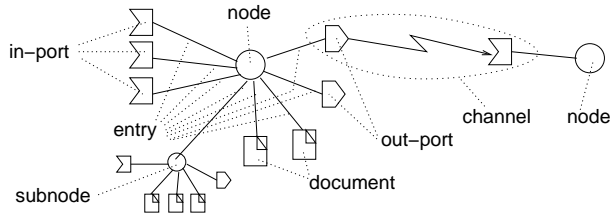}
    \caption{Detail of an example \ourModel network.}
    \label{fig:notation}
  \end{center}
\end{figure}

Our \ourModel framework captures the features and functionalities of
the many information system types that have emerged to support
individual requirements of information dissemination and retrieval. In
its role as an \emph{unifying} framework for heterogeneous
functionalities, it must provide the following features:

% \subsubsection{Features.}

\textit{Openness and extensibility:} our framework is generic in
the sense that support for arbitrary document and metadata formats, as
well as support for a wide range of query and filtering languages,
transmission protocols, and manipulations of document streams can be
plugged in.

\textit{Interoperability} with existing systems is a consequence of
the extensibility with various transmission protocols.

\textit{Protocol Transparency:} As Franklin and Zdonik put it in
\cite{DBIS-Framework}, \emph{``the character of a link should be of
concern only to the nodes on either end''}. This means that the model
treats different protocol characteristics such as push/pull or
periodic/aperiodic in a transparent way. Moreover, different and even
mixed models should be supported simultaneously as in the case of a
mailing-list server (push) with integrated archive (pull), or in the
case of an document repository (pull) offering a notification service
for its catalog (push).

\textit{Modularity} is achieved by offering a building kit of elements
that can be freely combined to build networks.

\textit{Evolution and Management:} IDNs must be able to adapt to
changing demand and supply of information by providing for dynamic
reconfiguration. The information dissemination mechanism lends itself
naturally to transport the necessary administrative
information. \ourModel ports are able to inspect and to change the
configuration of their parent nodes in reaction to events or external
requests.

\textit{Filtering, document conversion, (dis)aggregation:}
Ports may arbitrarily process the document streams passing through
them. This includes the deletion of selected documents (filtering),
arbitrary transformations of single documents, combination of
documents into aggregated documents (such as archives) as well as
disaggregation of documents into multiple documents.

\textit{Scheduling:} Many applications require the scheduled and
periodic delivery of information.  This is supported in our framework
by means of time events.

\textit{Coordination of Demand and Supply:} To support demand-driven
delivery of information, the information demand and its termination
must be propagated across a network from consumers to information
providers. The complementary requirement is the advertisement of
information offers to potential consumers.
 
\textit{Security}
is crucial in all distributed systems. Our model supports the
implementation of a large range of security policies. The use of
existing authentication and encryption protocols is supported through
its extensibility with various transmission protocols and the support
for arbitrary document conversions.

\subsubsection{Contents} 

The rest of the paper is organized as follows: we present the
conceptual model of \ourModel in Sec.~\ref{sec:data-model}. The
central protocol for communication between nodes and their ports is
presented in Sec.~\ref{sec:node/port-protocol}. Then we demonstrate in
Sec.~\ref{sec:applications} several applications of the \ourModel
approach. In Sec.~\ref{sec:related-work} we discuss related work.
\begin{TechnicalReport}
  Some extensions of our model are discussed in Sec.\ref{sec:extensions}.
\end{TechnicalReport}
We conclude and give an outlook in Sec.~\ref{sec:conclusion}.

\section{Conceptual Model}\label{sec:data-model}

\begin{figure}[htb]
  \begin{center}
    \includegraphics*[width=0.9\textwidth]{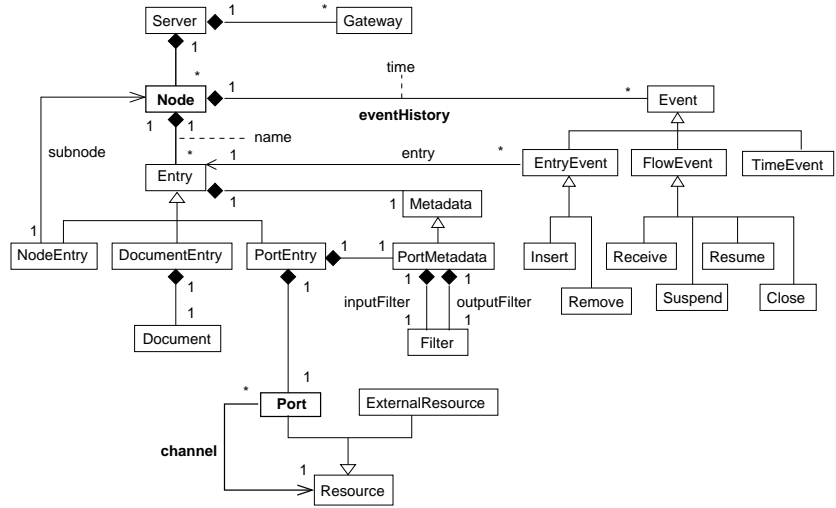}
    \caption{A conceptual model of \ourModel networks (UML class
      diagram).}
    \label{fig:schema}
  \end{center}
\end{figure}

% The elements of the \ourModel data model are described in detail in
% the next sections. In Fig.~\ref{fig:schema} the data model is
% presented in form of an UML class diagram. The principal entity types
% are \textsf{Document}, \textsf{Node}, and \textsf{Port}. Documents,
% ports, and subnodes are attached to nodes by means of
% \textsf{Entries}. 
%% <MYRA>
The principal components of the \ourModel framework are
\textbf{N}odes, \textbf{P}orts and \textbf{C}hannels. Intuitively,
documents are transferred between nodes across communication channels,
whereby a node may have more than one communication port. To capture
the functionalities of an IDN in a generic way, a more elaborate data
model is required, though, which is presented in Fig.~\ref{fig:schema}
as an UML class diagram. As can be seen in the left side of the
figure, a node may have documents, ports and subnodes (to be explained
later) attached to it by means of \textsf{Entries}.
%% </MYRA>
Nodes are contained within \textsf{Servers} that are located on
physical hosts.  \textsf{Channel}s are modeled as associations between
ports and \textsf{Resource}s which specialize into \textsf{Port}s and
\textsf{ExternalResource}s. The latter may be any external software
system offering a well-defined communication protocol; e.g., a Web
server/browser, a file system, a database server etc. A Port
communicates with its parent node by exchanging \textsf{Event}s (c.f.\
Sec.~\ref{sec:events},\ref{sec:node/port-protocol}) that are archived
in an \textsf{eventHistory}. It communicates with resources on other
servers via \textsf{Gateway}s (c.f.\ Sec.~\ref{sec:gateways}). We
stress that each node, port, and gateway in an \ourModel network may
have an individual implementation which enables the support for
heterogeneous document and metadata formats, protocols, conversions,
and filter languages.

\begin{example}{\emph{(Forwarding)}}\label{example:forwarding} 
  A simple \ourModel application forwards all new documents from a
  ``\emph{provider node}'' to a ``\emph{consumer node}'' via a
  channel. A new document entry is inserted by posting an \textsf{Insert}
  event to the provider node. A ``\emph{provider port}'' at this node
  consumes the \textsf{Insert} event and communicates its contents to
  a ``\emph{consumer port}'' at the consumer node (via suitable
  gateway objects understanding a common communication protocol). The
  consumer port inserts the document by emitting a corresponding
  \textsf{Insert} event. In addition, the provider port may delete
  forwarded document entries by emitting \textsf{Remove} events for
  them.
\end{example}

\subsection{Nodes}\label{sec:nodes}

Each \textsf{Node} stores a set of \textsf{Entries} that are uniquely
named within the node. Each entry belongs to a single node and
consists of: (i) a metadata record \emph{describing} the content of
the entry, (ii) the proper content of the entry, which may be either a
\textsf{Document}, a \textsf{Node}, or a \textsf{Port}.  Hence we
speak of \emph{document entries}, \emph{node entries}, and \emph{port
entries}.
Node entries enable the construction of node hierarchies similar to
directory trees in file systems.  We require that each server has a
unique root node.  Then each entry can be uniquely addressed by a URL
that consists of the host address of the server and a sequence of
entry names describing the path from the root node via a chain of
subnodes to the entry. 
% This is analogous to URLs for conventional Web sites.

\subsection{Events}\label{sec:events}

\textsf{Event}s describe state changes in a node. The
state of a node is its content, i.e., its set of entries. 
% The state
% space of a port depends on its implementation. For instance, an
% out-port may have the states \emph{ready/not ready} (to receive the
% next event).  
Each node keeps a history of events ordered by creation time.  Every
event posted by a port is received by its node, recorded in the event
history, and delivered to all other ports. The two most basic types of
events produced by ports are \textsf{Insert} and \textsf{Remove}
events. An \textsf{Insert} event causes the node to insert a new entry
specified by the event. Similarly, a \textsf{Remove} event causes the
node to delete the entry named by the event. Newer events may cancel
older events in the event history. For instance, a \textsf{Remove} event
for an entry may delete the preceding \textsf{Insert} event for this
entry. In addition, a node may have some policy for discarding old
events.

\begin{example}{\emph{(Replication)}}\label{example:replication}
  Example~\ref{example:forwarding} can be extended to support
  mirroring of repositories by transmitting not only \textsf{Insert}
  events, but also \textsf{Remove} events. Thus both insertions and
  deletions at the provider node are propagated to the consumer node
  where they are mirrored.
  By adding a back channel, bidirectional mirroring can be
  achieved. To distinguish originals from replicas each entry is
  tagged with its originating node. More complex replication schemes
  such as the I2-DSI architecture\cite{I2-DSI} can be supported as
  well.
\end{example}

Entries can be accessed through the events referring to them. In
particular, the contents of a removed entry may be still accessible
through the event describing its removal, depending on the node
implementation. In addition to \textsf{EntryEvent}s referencing an
entry, there are further event types such as \textsf{FlowEvent}s for
flow control and \textsf{TimeEvent}s for scheduling
(c.f. Fig.~\ref{fig:schema}). 
% Each time event cancels all former time events in the event history.

\subsection{Gateways}\label{sec:gateways}

\begin{figure}[htb]
  \begin{center}
    \includegraphics*[height=0.125\textheight]{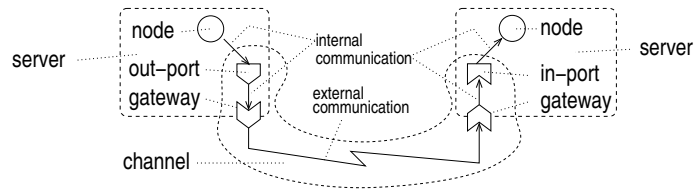}
    \caption{Implementation of a channel.}
    \label{fig:channel}
  \end{center}
\end{figure}

Each server hosts a number of \textsf{Gateway}s which are in charge of
communication with (gateways at) other servers or external
resources. Each gateway implements a specific communication
protocol. The communication between the ports forming a channel is
routed via gateways as illustrated in Fig.~\ref{fig:channel}. Gateways
and ports communicate via a protocol-specific API. Incoming
connections may indicate to the gateway which port they want
to contact by specifying its URL.

\subsection{Ports}\label{sec:ports}

Ports are persistent processes that enable the communication between
nodes and resources, including the ports of other nodes, via
gateways. Most ports are unidirectional in- and out-ports. In-ports
transform an incoming data stream into a stream of events that are
posted to the parent node. Conversely, out-ports consume an event
stream and turn it into an outgoing data stream.
% \footnote{Additional
% time and flow-control events may go up-stream, cf.\
% Sec.~\ref{sec:node/port-protocol}}.

\begin{example}{\emph{(Filtering, Conversion, Aggregation})}
  The ports involved in Example~\ref{example:replication} may be
  designed to forward/accept only documents matching certain filtering
  criteria. The ports may convert the forwarded documents, for
  instance by applying XSL stylesheets to XML documents.  The provider
  port may even pack several documents into archive files which are
  unpacked by the consumer port.
\end{example}

\subsubsection{Properties of a Port:}
(i) Ports may be in-ports, out-ports, or bidirectional. (ii) Their
operation may be periodic or a-periodic. (iii) The communication with
the external resource may follow the push-, pull-, or a mixed model.
We call the combinations in-port + push and out-port + pull
\emph{externally driven}, the combinations in-port + pull and out-port
+ push \emph{internally driven}.

\begin{example}{\emph{(Push vs.\ Pull)}}
  Example \ref{example:forwarding} can be implemented using either
  push- or pull-mode communication. In \emph{aperiodic push-mode}, the
  provider port is triggered by the \textsf{Insert} events posted to
  the event history of the provider node. For each \textsf{Insert}
  event the provider port feeds the new document to the consumer port.
  In \emph{periodic pull-mode}, the consumer port is triggered by time
  events at the consumer node to poll the provider port for new
  documents. Only if triggered by the consumer port does the provider
  port request the next insert events from the provider node and
  transmit the referred documents to the consumer port. An application
  of this scheme would be to periodically fetch new metadata records
  from an OAI\cite{OAI} repository.
\end{example}

\subsubsection{Event filters.}
For security and efficiency reasons, ports need not receive all events
nor should they be able to emit arbitrary events. Each port entry
contains as metadata two filter expressions, the \emph{input} and the
\emph{output filter}. They are interpreted by the node to restrict
which events the port may receive and produce, respectively. Proper
in-/out-ports can be enforced by blocking output/input filters.
Different node implementations may support different filter languages
of varying expressiveness.  Filter conditions may reference properties
of events as well as metadata and content of the entry an
\textsf{EntryEvent} refers to.

\textit{Metaports.} Ports that are restricted to receive or generate
\textsf{Insert} or \textsf{Remove} events \emph{for document entries
only} are called \emph{plain ports}. All other ports we call
\emph{metaports}. Only metaports are allowed to inspect, insert, or
delete port entries and subnode entries. Metaports provide a reflexive
and introspective mechanism for evolution and remote management of
\ourModel networks. Metaports are called \emph{admin ports} if their
filters allow arbitrary events to pass. For security reasons,
communication with metaports (and admin ports in particular) should be
protected by suitable authentication and encryption techniques.

\textit{Security Policies} have to be enforced by choosing appropriate
input/output filters. Basic objectives of such a policy are to ensure
that plain ports can only read or write document entries and that meta
ports cannot insert port entries which violate the security
policy. Further objectives may be to hide certain entries from certain
ports or to restrict which ports can delete certain entries.

% \textit{Admin ports.} Typically a node has at least one
% metaport which is used for administering the node.  The filters of
% such an \emph{admin port} do not block any events in order to allow
% arbitrary manipulations of the node content. Hence external access to
% this port should be restricted using appropriate authentication and
% encryption mechanisms.

\subsection{Channels}\label{sec:channels}

\ourModel \textsf{channel}s are unidirectional communication
relationships between nodes. A channel transmits a stream of data
items from a \emph{source node} to a \emph{target node} via a port and
a gateway at each end of the channel (c.f. Fig.~\ref{fig:channel}). A
channel may also connect a node to some other external
resource. Bidirectional streams between nodes are implemented as pairs
of channels.

% Channels can be viewed at different levels of abstraction:
\emph{Document channel}s transmit streams of documents. Document
channels are generalized by \emph{entry channels} that copy entries
from the source node to the target node. Entry channels allow to
transmit also non-document entries. \emph{Event channels} generalize
entry channels further by transmitting arbitrary event streams which
may also propagate deletions of entries. 

% Every event channel is implemented by a pair of ports, i.e., by an
% out-port at the source node and an in-port at the target node. Note,
% that ports (and hence channels) allow arbitrary filtering,
% transformation, aggregation, and disaggregation of the transmitted
% events. This property carries over to entry channels and document
% channels.

Ports may be created without being part of a channel and they may
survive the termination of the channel they are participating
in. Ports are intended to participate in no more than one channel at a
time, except in cases like bidirectional ports. Instead of multiple
channels ending in the same port, the advisable method is to create a
dedicated port for each new channel.

To establish a new channel between a source node and a target node, a
source port and a target port need to exist. One of these ports
contacts the other port (via suitable gateways).
% , either triggered by
% an event received from its node, or by its own initiative.
Ports may delegate incoming connections to other ports and may even
create new ports to this purpose. Thus a node can support arbitrary
numbers of incoming or outgoing channels with dedicated ports. Various
policies for connection pools can be implemented as well this way.

\section{Unifying the Push- and Pull-Model}\label{sec:node/port-protocol}

\begin{TechnicalReport}
  
  The communication between a node and its ports via the event history
  is illustrated in Fig.~\ref{fig:event-history}. This diagram depicts
  an in-port that currently deletes \textsf{entry 1} by writing a
  \emph{delete event} and an out-port that is just reading an earlier
  event describing the insertion of \textsf{entry 2}. Both ports are
  communicating with external resources. The details of the
  interaction between a port and the event queue are specified by the
  node/port protocol that is introduced in the sequel of this section.

  \begin{figure}[htb]
    \begin{center}
      \includegraphics*[scale=0.667]{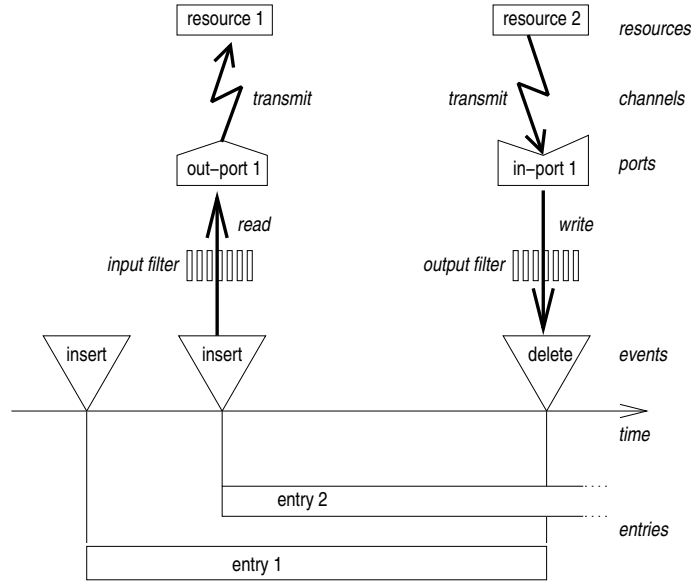}
      \caption{The event history and ports of a node.}
      \label{fig:event-history}
    \end{center}
  \end{figure}
  
\end{TechnicalReport}

As argued in the introduction, information dissemination frameworks
need to support various communication models. In the \ourModel
framework, both push-model and pull-model communication are
generalized into the node/port communication protocol. The
communication between a node and its ports is entirely event-based.  A
port can only post events to its parent node that pass its output
filter. On arrival at the node, events may trigger state changes such
as insertion or removal of entries. By default, every event is entered
into the node's event history. \textsf{FlowEvent}s control the
delivery of events to a port.

\begin{figure}[htb]
  \begin{center}
    \includegraphics*[height=0.2\textheight]{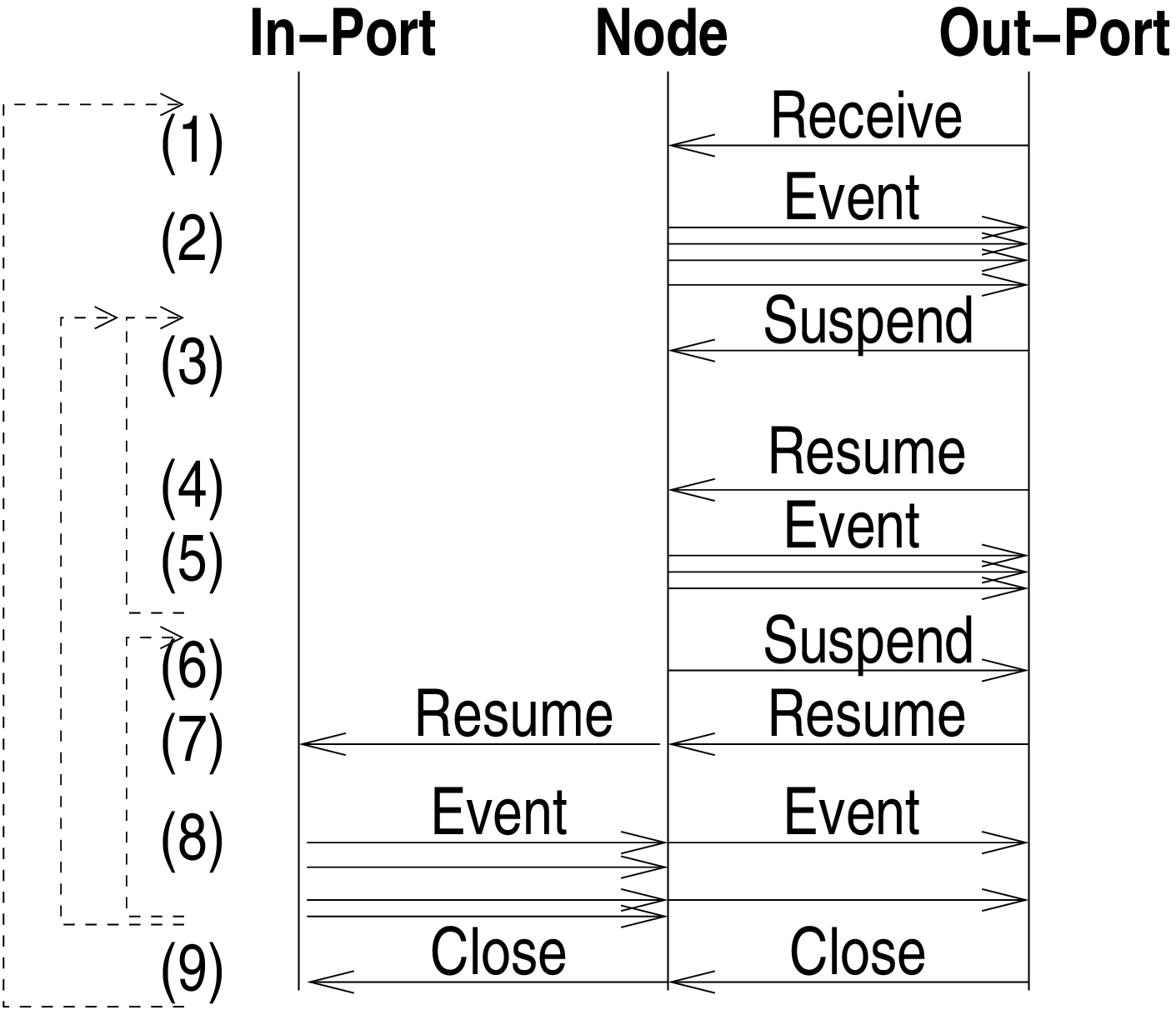}
    \caption{The node/port protocol as an UML sequence diagram.}
    \label{fig:node-port-protocol}
  \end{center}
\end{figure}

The sequence diagram in Fig.~\ref{fig:node-port-protocol} illustrates
the flow control in the node/port protocol.  Each port entry stores a
\textsf{receive} flag to indicate whether the port is currently
accepting input events. The port may set this flag by issuing a
\textsf{Receive} event which is not added to the event history
(1). After the \textsf{receive} flag has been set, the node starts
delivering events from the event history to the port (2). Only events
passing the input filter of the port entry are delivered to this
port. Each port entry also contains a \textsf{cursor} pointing into
the event history to keep track of the events already delivered. The
node implementation should guarantee a fair delivery of events, i.e.,
eventually every interested port will receive every event. Nodes may
also implement delivery policies that honor \emph{Quality-of-Service}
requirements stated in the metadata of port entries.

A port may suspend the delivery of events temporarily by issuing a
\textsf{Suspend} event which clears its \textsf{receive} flag
(3). After having issued a \textsf{Resume} event (4) the
\textsf{receive} flag is set again and the port continues to receive
events (5).  After the cursor of a port has reached the end of the
event history, the node issues a \textsf{Suspend} event to the port
which clears its \textsf{receive} flag (6). If the port wants to wait
for future events, it has to emit again a \textsf{Resume} event
(7). If the cursor is still at the end of the event history at this
time, the \textsf{Resume} event is posted to the event history which
may trigger interested in-ports to post events matching the
input-filter of the out-port (8). Only those events that actually pass
the input-filter are delivered to the out-port.  A port can terminate
the delivery of events by issuing a \textsf{Close} event (9). This
cancels any preceding \textsf{Resume} events and is received by
interested in-ports which in turn may stop event delivery.

An important advantage of the node/port-protocol is its support both
for the push and pull communication model:
In the \textit{push-model} external producers send data to in-ports
of a node which post it in form of \textsf{Insert} events. Out-ports
of this node receive these events and forward the data to external
consumers, including in-ports of other nodes.  Thus information can be
pushed from providers to consumers across a \ourModel network.

The \textit{pull-model} requires more coordination: An external
consumer connected to an out-port causes it to emit a \textsf{Receive}
event. After having received all historic events the out-port issues a
\textsf{Resume} event which is received by all interested in-ports
(c.f.\ Fig.~\ref{fig:node-port-protocol}). These in-ports may retrieve
data from external producers and post it in form of \textsf{Insert}
events that are received by all out-ports with matching input filters,
including the out-port that has triggered the import. This enables
consumers to pull information from providers across a \ourModel
network. In-ports may achieve finer flow control by also monitoring
\textsf{Suspend} and \textsf{Close} events.

\section{Applications}\label{sec:applications}

In this section we present some example applications and how they can
be implemented using the \ourModel approach. 

% \subsection{Replication}\label{sec:mirroring}

% The document content of a ``source'' node can be replicated in some
% other ``target'' node via a channel: at the source node, an out-port
% communicates all \emph{insert} and \emph{remove events} for document
% entries to an in-port at the target node which posts them to the
% target node.  Bidirectional replication is enabled by adding another
% channel in the reverse direction. A ping-pong effect can be avoided by
% tagging the metadata of each entry with its origin. More complex
% replication schemes such as the I2-DSI architecture\cite{I2-DSI} can
% be supported as well.

\begin{figure}[htbp]
  \begin{center}
    \includegraphics*[width=0.5\textwidth]{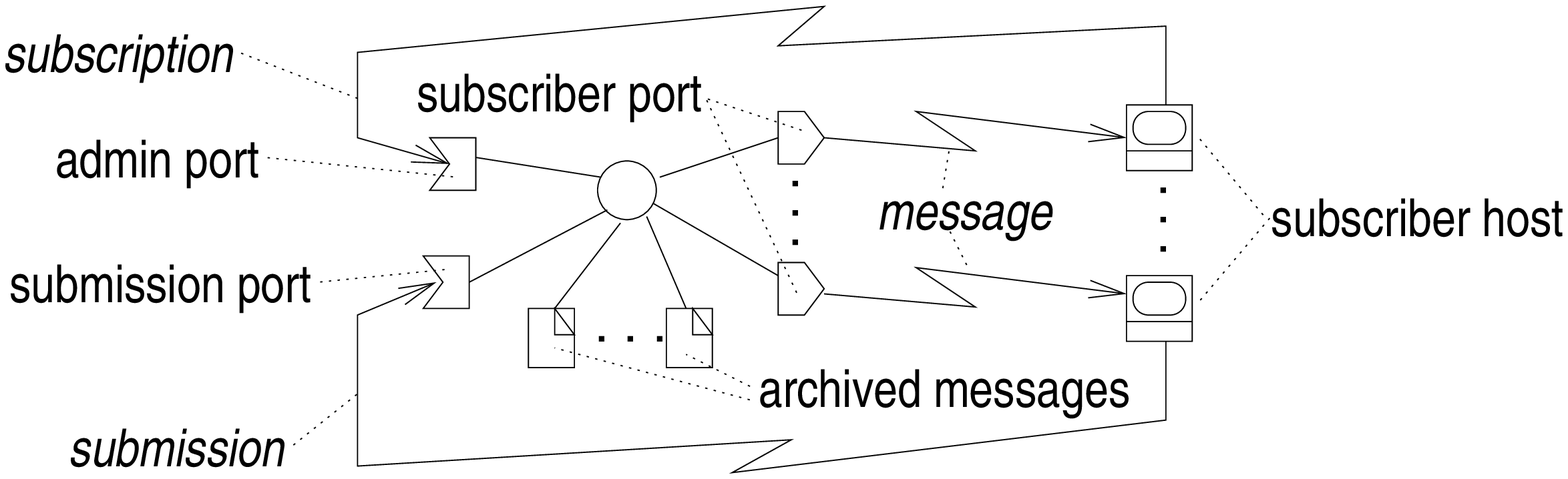}
    \caption{A mailing list (gateways and mail server omitted).}
    \label{fig:mailinglist}
  \end{center}
\end{figure}

% \subsubsection{Mailing Lists.}\label{sec:mailing-list}
\mysubsubsection{Mailing Lists.}\label{sec:mailing-list}
A mailing list can be implemented using a single node
(Fig.~\ref{fig:mailinglist}). An active ``administration'' in-port
periodically fetches (un-)subscription messages via some gateway from
an external mail server. The administrator port transforms each
(un-)subscription message into an \emph{insert (remove) event} of a
``subscriber'' out-port.  A ``submission port'' periodically fetches
submissions from the mail server. For each received message it emits
an \textsf{Insert} event which is received by the subscriber ports and
sent to each subscriber using a gateway to the external mail
server. Subscriber ports may convert messages into a format preferred
by the subscriber and bundle several messages into message digests.

All messages are archived in the mailing list node (unless a ``garbage
collector'' port is deleting old messages). In paragraph ``Simple
Document Server'' below we explain how to add a Web interface for this
archive.

% The administration port may be extended to understand messages
% containing requests for message digests. Each digest request is
% converted into a ``digest'' sink port that scans the existing
% messages, selects those requested, packs them into a single message,
% and sends this digest to the sender of the request, after which the
% digest port terminates and is removed from the node.

% A mailing list server supporting the dynamic creation of mailing lists
% can be implemented by a root node. This root node has an
% administrative port that inserts for each incoming request for a
% mailing list a new mailing list node (as described above) as a
% subnode.

% Moderated mailing lists can be implemented by introducing a second
% node having a new public submit port to which all messages are
% sent. This node forwards all messages to the moderator. The moderator
% sends all approved messages to the submit port of the original mailing
% list node. The submit port of this node is turned into a private port
% that accepts only mails signed by the moderator.

%\subsubsection{Alerting Systems.}\label{sec:alerting-systems}
\mysubsubsection{Alerting Systems.}\label{sec:alerting-systems}
The mailing list application can be easily extended to build alerting
systems that filter the incoming documents on behalf of their
subscribers. Each subscriber port uses its input filter (and possibly
internal post-processing) to forward only those documents the
subscriber is interested in. To handle a large number of subscriptions
efficiently, the node may index the input filters of its subscriber
ports to deliver incoming events only to matching subscriber ports
(c.f.\ \cite{YGM94}).

\begin{TechnicalReport}
                                
% \subsubsection{Hierarchical Document Classification.}\label{sec:document-filtering}
\mysubsubsection{Hierarchical Document Classification.}\label{sec:document-filtering}
Documents can be classified into a subject hierarchy by modeling the
subject hierarchy as an \ourModel network as depicted in
Fig.~\ref{fig:filtering}. Each node represents a subject class, its
subnodes represent its subclasses. For each subnode there exists an
out-port that forwards messages to an in-port of the subnode. This
out-port implements a classifier that permits only those documents
belonging to the class represented by the subnode. Note that
classifiers need not necessarily be automated, they may as well prompt
a human expert for advice.

\begin{figure}[htbp]
  \begin{center}
    \includegraphics*[height=0.2\textheight]{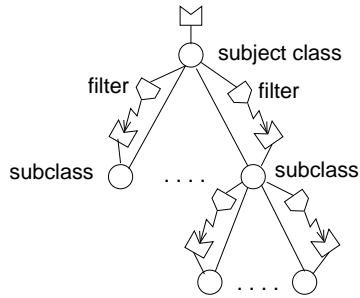}
    \caption{A classification hierarchy for document filtering.}
    \label{fig:filtering}
  \end{center}
\end{figure}

Incoming documents are fed into the root node of this hierarchy
and are distributed to those nodes representing the classes they
fit. 
% Different classification policies can be implemented: 
% \begin{itemize}
% \item documents can be kept or deleted in a subject class after
%   they have been forwarded to subclasses.
% \item documents can be assigned to the first matching subclass only by
% using the extension described in Sec.~\ref{sec:exclusive-consumption}.
% \end{itemize}

\end{TechnicalReport}

% \subsubsection{Simple Document Server.}\label{sec:document-server}
\mysubsubsection{Simple Document Server.}\label{sec:document-server}
A node storing a document collection can act as a simple (HTTP)
document server. A document request from a client is received by an
admin-port via a HTTP gateway which inserts a dedicated ``request
port'' into the node. The request port specifies the name of the
requested document in its input filter. After receiving the recorded
\textsf{Insert} event for the requested document entry, the request port
reads the document stored in this entry and sends it to the
client. For a missing document the request port receives a
\textsf{Suspend} event instead and sends an error message to the
client. In both cases, the port removes itself from the node by
emitting a \textsf{Remove} event for itself.

% \subsubsection{Information Retrieval.}\label{sec:information-retrieval}
\mysubsubsection{Information Retrieval.}\label{sec:information-retrieval}
Information retrieval from a document repository is a generalization
of the simple document server application mentioned above.
%  from Section~\ref{sec:document-server}. 
Instead of request ports we speak of ``query ports'' in this
context. Instead of a single document, a query port may return a set
of documents.

The input filter of a query port preselects potential result
documents. The query port post-processes the events passed by the
filter depending on the expressiveness of the node's filter
language. Each matching document is returned to the client. A node may
maintain an internal index for its event history and use it when
evaluating new filter expressions against existing events.
% It is important that the filter language is
% expressive enough to effectively reduce the amount of post-filtering
% that is left to the ports since the latter cannot be supported by
% indices.

The query port may also collect and order the matching documents
according to some relevance measure. The query node receives a
\textsf{Suspend} event after the event history is exhausted. It then
delivers the result as a ranked list to the client either in one part
or in several portions of fixed size.

% \subsubsection{Query Mediation.}\label{sec:query-mediation}
\mysubsubsection{Query Mediation.}\label{sec:query-mediation} In
networked information retrieval, a query from a client against a
mediator is processed by forwarding it to several source
repositories. Wrappers translate between the mediator query language
and the heterogeneous source query languages. They also convert the
returned results into a common format. The mediator returns the
combined results to the client.

In \ourModel, a query mediator can be implemented as follows: Each
query port entry states its query in its metadata. For each source
repository there exists a ``query translator port'' in the mediator
node which monitors the \textsf{Receive} events posted by query ports,
analyzes each query and decides whether to delegate it to its source
repository. In this case it inserts a ``wrapper port'' supplied with
the name of the query port repository-specific translation of the
query. The query port name is also stored in the wrapper port
metadata. The wrapper port queries its source repository, posts the
received results (tagged with the query port name), and finally
removes itself from the node.

The query port subscribes to \textsf{Insert} events for document
entries tagged with its own name, removes them from the node, and
returns the documents to the client. It may also filter, collect,
sort, and group the results before returning them. The query port also
monitors insertions and removals of wrapper ports to check whether all
wrapper ports for the query have terminated.

If the client aborts the query, the query port removes all wrapper
ports for this query. Each wrapper port is notified of its own removal
by a \textsf{Remove} event which triggers it to propagate the abort to
its repository and then to terminate. Finally, the query port removes
itself and terminates as well.
% \end{TechnicalReport}

% \subsubsection{Further applications.}\label{sec:other-applications}
\mysubsubsection{Further application
areas}\label{sec:other-applications} include workflow management
(using a document-flow centered view), hierarchical document
filtering, network management, Web-caching, and peer-to-peer
information dissemination networks.

\begin{TechnicalReport}
  \section{Extensions}\label{sec:extensions}

  \subsection{Exclusive consumption of events} \label{sec:exclusive-consumption}

  In certain applications it is important to deliver an event
  exclusively to a single port. When a node receives an event, it
  returns a flag that determines whether the event is consumed by the
  port or delivered to further ports. By attaching a priority to each
  port, a partial ordering for the event delivery is introduced. Ports
  of equal priority may be ordered in a nondeterministic way.

% When an
% event is delivered to a port, the port acknowledges the event with a
% boolean value. This value indicates whether the event is to be
% delivered to any further ports of the same or of lower priority. The
% event is tagged with the priority of this port, but kept in the event
% queue. Thus a port inserted later will still be able to receive the
% event if its priority is sufficiently high. The only way to prevent an
% event to be delivered to a high-priority port that is inserted later
% is to cancel this event by emitting an appropriate complementary event
% as mentioned above.

  \subsection{External Security Policies}\label{sec:firewall}

  In the \ourModel model presented so far, every port is free to
  communicate with arbitrary resources via suitable gateway
  objects. Input and output filters control only the information
  exchange between nodes and ports. The right place to control the
  communication between ports and \emph{external} resources or ports
  is at the gateway objects.

  External security policies which restrict this communication may be
  defined as sets of \emph{firewall rules} similar to those used by
  typical firewall software. Such security policies may be either
  global or are attached as metadata to port or subnode entries.  A
  security policy attached to a subnode entry applies to all ports in
  the tree rooted in the subnode entry and is subject to refinement by
  security policies in nodes or ports deeper within this tree.

  This approach can be extended to control intra-server communication
  by requiring that all communication between ports on the same server
  has to be routed as well through gateway objects.

  \subsection{Dynamic subscriptions}\label{sec:multi-subscriptions}

  The input filter of a port serves a twofold purpose: it reflects the
  security policy of the node and the interests of the port. While the
  security policy can be considered to be static, the interests of the
  port may change over time. Moreover, a port may want to maintain
  several subscriptions for the event queue that it can control
  independently (via \emph{receive}, \emph{suspend}, \emph{resume},
  \emph{close} events).

  For instance, a port may initially subscribe to a future time event
  and only after receiving this time event it will subscribe to other
  events.

  To cater these needs we propose to change the node/port protocol as
  follows: two parameters are added to \emph{receive events}, a filter
  expression and a subscription name chosen by the port. The
  subscription name is used to reference the subscription in
  \emph{suspend}, \emph{resume}, and \emph{close} events. Subscription
  names need to be unique only within each port.

  In the example, the port would first emit a \emph{receive} event
  specifying a filter condition for a future time event. After
  receiving this time event, the port would emit a \emph{close} event
  for this first subscription and a \emph{receive} event specifying
  the second subscription.

\end{TechnicalReport}

\section{Related Work}\label{sec:related-work}

\textit{Message-oriented middleware} such as the Java Messaging
  Service (JMS) standard \cite{jms99}, IBM MQ-Series \cite{MQ-Series},
Oracle Advanced Queueing \cite{OracleAQ}, and Talarian SmartSockets
\cite{SmartSockets} offer message queues and topics to connect
supplier and consumer clients. Extra client-side programming is needed
to connect queues or topics to a larger network. Clients may specify
filter-expressions on the message header fields in a simple and fixed
filter language. The object-based \textit{CORBA notification service}
\cite{CNS} does not have these shortcomings. Moreover it supports the
propagation of demand for and the supply of certain event
types. However, this mechanism is external to the event propagation
mechanism and it does not allow the propagation of more complex
queries as in \ourModel. All mentioned systems are missing
introspective means for management and evolution as in our model.
Channels in our model could be based on these systems.

\textit{Siena}~\cite{Siena01} is a push-model publish/subscribe system
for alerting within a wide-area network. It offers scalability by
distributing filters over servers within the network and saving
bandwidth by filtering close to the event sources and bundling similar
subscriptions. Siena is modular and offers sophisticated filtering
mechanisms including dynamic configuration and distribution. It lacks
openness, document stream transformation, and scheduling.

\textit{Elvin4} \cite{Elvin4} is an alerting system based on a
client-server-client architecture. It is open for different transport,
security, and marshalling mechanisms. Elvin4 is not intended to scale
to wide-area networks or to go beyond publish/subscribe systems. Its
\emph{quenching} mechanism prevents the publication of events for
which there exists no consumer. This is similar, but less general than
the query-propagation mechanism offered by our model.

The \textit{Information and Content Exchange (ICE) protocol}
\cite{ICE1.1} is an XML- and HTTP-based standard for
content syndication that could be supported by our
framework. Providers advertise information delivery offers
which consumers may subscribe to. Both push- and pull-delivery is
supported. 

Alerting systems based on \textit{Continual Queries} such as CQ
\cite{liu98,liu99} and WebCQ \cite{WebCQ02} have been developed to
track changes in databases and on Web sites. The focus here is on
efficient detection and meaningful summarization rather than on
openness or modularity. CQ systems could be used as information
providers within the \ourModel framework.

For a survey on further event-based systems, see \cite{rifkin98a}.

\section{Conclusion and Outlook}\label{sec:conclusion}

The \ourModel approach presented here offers an unifying framework for
building a wide range of information dissemination networks 
% that can be composed in a modular way 
built from nodes, ports, and channels.
% Its main advantages compared to existing IDs are its
% modularity, openness, interoperability. Due to its openness, existing
% filtering, transformation and aggregation/disaggregation technology
% can be easily integrated into nodes and ports.  Other important
% features are reflexive configurability and demand propagation through
% networks.
It is open for heterogeneous protocols, filtering languages, document
and metadata formats by allowing different node and port
implementations within the same IDN. Moreover, existing document
processing technology can be easily incorporated into nodes and ports
which enables all kinds of filtering operations, format conversions,
and (dis)aggregation operations. Due to its openness, \ourModel is
also interoperable with a wide range of existing IDS. By using a
general event-based node/port communication protocol, the
heterogeneity of the underlying port/port communication protocols is
hidden. In particular, it supports both supply-driven (push-model) and
demand-driven (pull-model) information dissemination
%  and the propagation of information demand 
across an IDN. Other important features are dynamic and reflexive
configurability, support of scheduling, and of security policies.
%  based on input/output filters.

The generality of the \ourModel approach supports various
applications, from newsgroups to query mediation, and facilitates the
development of innovative information dissemination applications.
 
% We envisage that the openness of the \ourModel framework will also
% facilitate its implementation since one can take advantage of existing
% technology such as for instance HTTP servers or message-oriented
% middleware.

Due to space limitations several interesting extensions had to be
omitted, e.g., complex events, exclusive consumption of events,
dynamic subscriptions, and external security policies. A Java
implementation of the node/port protocol is available at
\cite{NPCHomePage}. We are currently extending it to a full-fledged
IDN construction kit.

\bibliographystyle{alpha}
\bibliography{npc}

\end{document}